\documentclass[twocolumn]{aastex631}
\usepackage{multirow}
\usepackage{CJK}

\newcommand{\obj}{NGC~3628}

\newcommand{\softw}{\textsc}
\newcommand{\module}{\texttt}
\newcommand{\telescope}{\textit}
\newcommand{\Msun}{\ifmmode {$M$_{\odot}}\else{M$_{\odot}$}\fi}
\newcommand{\hii}{H\,\textsc{ii}}

\received{XXX}
\revised{XXX}
\accepted{XXX}

\submitjournal{ApJ}

\shorttitle{SNRs in \obj}
\shortauthors{Yang \& Ou}
\graphicspath{{./}{figures/}}

\begin{document}


\title{The core starbursts of the galaxy \obj: Radio very long baseline interferometry and X-ray studies}

\begin{CJK*}{UTF8}{gbsn}
\correspondingauthor{Xiaolong Yang; Ziwei Ou}
\email{yangxl@shao.ac.cn; ouzw3@mail.sysu.edu.cn}

\author[0000-0002-4439-5580]{Xiaolong Yang (杨小龙)}
\affiliation{Shanghai Astronomical Observatory, Chinese Academy of Sciences, Shanghai 200030, China}
\affiliation{Shanghai Key Laboratory of Space Navigation and Positioning Techniques, Shanghai 200030, China}

\author[0000-0002-3632-474X]{Ziwei Ou (区子维)}
\affiliation{School of Physics and Astronomy, Sun Yat-sen University, Zhuhai 519082, China}



\begin{abstract}
We present radio very long baseline interferometry (VLBI) and X-ray studies of the starburst galaxy \obj. The VLBI observation at 1.5\,GHz reveals seven compact (0.7$-$7\,parsec) radio sources in the central $\sim$250\,parsec region of \obj. Based on their morphology, high radio brightness temperatures ($10^5-10^7$\,K), and steep radio spectra, none of these seven sources can be associated with active galactic nuclei (AGNs); instead, they can be identified as supernova remnants (SNRs), with three of them appearing consistent with partial shells. Notably, one of them (W2) is likely a nascent radio supernova and appears to be consistent with the star formation rate of \obj\ when assuming a canonical initial mass function. The VLBI observation provides the first precise measurement of the diameter of the radio sources in \obj, which allow us to fit a well-constrained radio surface brightness - diameter ($\Sigma-D$) correlation by including the detected SNRs. Furthermore, future VLBI observations can be conducted to measure the expansion velocity of the detected SNRs. In addition to our radio VLBI study, we analyze \telescope{Chandra} and \telescope{XMM-Newton} spectra of \obj. The spectral fitting indicates that the SNR activities could well account for the observed X-ray emissions. Along with the \telescope{Chandra} X-ray image, it further reveals that the X-ray emission is likely maintained by the galactic-scale outflow triggered by SN activities. These results provide strong evidence that SN-triggered activities play a critical role in generating both radio and X-ray emissions in \obj\ and further suggest that \obj\ is in an early stage of starbursts. 
\end{abstract}



\section{Introduction\label{sec:intro}}

Galaxies with enhanced star formation compared with normal spiral galaxies, especially in the central nuclear region, are known as starburst galaxies \citep{1996ApJ...472..546L}. They have star formation rates (SFRs) several to hundreds of times higher than ordinary galaxies like the Milky Way. Studying nearby starburst galaxies may shed light on our understanding of the evolution of their high redshift cousin in the early Universe \citep[e.g.][]{2022ApJ...930..128R}.

Starburst is an important episode in galaxy evolution \citep{2006ApJS..163....1H, 2018Natur.558..260Z}; it lasts up to a few hundred million years \citep[e.g.][]{2018MNRAS.477.3164M}. Simulations and observations indicate that the starburst activity can be triggered by mergers or close flybys between two galaxies \citep{1996ApJ...464..641M, 2005ASSL..329..143S, 2007A&A...468...61D, 2010A&A...518A..56M}. These interactions push the gas inflow toward the center of galaxies, where it reaches sufficient density to cause large numbers of massive stars to form \citep{2000ASPC..197...15C}. In addition, the birth (stellar feedback) and death (supernova explosion) of massive stars can generate shocks to compress the gas and further enhance the star formation \citep{1987ApJ...317..190M, 2008ApJ...684..978S, 2010ApJ...710L.142F, 2015ApJ...809..187S}. However, the interplay between a galactic scale outflow and the gas inflow and compression will regulate, slow down and eventually brake the intense star-forming process \citep{1985Natur.317...44C, 1988ApJ...330..695T,1990ApJS...74..833H, 2008ApJ...674..157C}.

Star-forming galaxies typically host rich populations of ultradense (UD) \hii\ regions \citep{1999ApJ...527..154K, 2001ApJ...559..864J} where massive stars are born. On the other hand, the large population of massive stars will rapidly evolve and end their lifetime with supernova explosions (supernovae, SNe) and form supernova remnants (SNRs) \citep{1998ApJ...505..639E}. These UD \hii\ regions are probes of recent star formation, while SNe and SNRs are vital tracers of star formation history and are responsible for the galactic scale feedback. The central regions of the majority of starbursts are obscured optically. However, radio emissions, which are unaffected by extinction, allow the central regions of these galaxies to be probed. The radio emissions include both non-thermal (synchrotron) emission, arising from the acceleration of particles to relativistic speeds in SNe and SNRs, and thermal free-free emission from UD \hii\ regions \citep{1992ARA&A..30..575C}. Furthermore, X-ray binaries (XRBs) and active galactic nuclei (AGNs) may also reside in star-forming galaxies, especially, AGNs can produce both thermal \citep{2004ApJ...613..794G, 2008MNRAS.390..847L, 2016MNRAS.459.2082R, 2018ApJ...869..114I, 2021MNRAS.508..680B, 2022MNRAS.514.6215Y} and non-thermal \citep{2014MNRAS.442..784Z, 2015MNRAS.447.3612N, 2020ApJ...904..200Y, 2022arXiv221100050Y, 2022MNRAS.517.4959Y} radio emissions; they may also contribute significant radio emissions in starburst galaxies \citep[e.g.][]{2004ApJ...613..794G, 2007MNRAS.377..731M}.

High-resolution very long baseline interferometry (VLBI) observations can resolve (constrain the size of) individual UD \hii\ region, SN, and SNR in nearby star-forming galaxies. The great success of using VLBI on this subject focuses on several nearby star-forming galaxies, e.g. M82 \citep{1999MNRAS.307..761P, 2006MNRAS.369.1221B, 2010MNRAS.408..607F}, Arp\,220 \citep{2005MNRAS.359..827R, 2006ApJ...647..185L, 2007ApJ...659..314P, 2011ApJ...740...95B}, Arp\,299 \citep{2004ApJ...611..186N, 2009AJ....138.1529U, 2012A&A...539A.134B}, NGC\,253 \citep{2014AJ....147....5R} and NGC\,4945 \citep{2009AJ....137..537L}. Observations of these objects may help us to understand the environment of starburst cores \citep{2011ApJ...740...95B}, since the evolution of SNe and SNRs is regulated by dense interstellar medium (ISM). Furthermore, SNe and SNRs are likely responsible for the origin of the well-known FIR-radio correlation and the galactic scale outflow, and the VLBI observations of these sources may provide useful inputs for models. Especially, with VLBI monitoring observations, it is possible to track the evolution of individual SN and SNR, and alert nascent SNe \citep{1995Sci...270.1475M, 2009A&A...505..927M, 2019A&A...623A.173V}. In this work, we will visit the starburst galaxy \obj.

\obj\ is a nearby \citep[$D = 10.6\,\mathrm{Mpc}$ or $z=0.0028$,][]{2015MNRAS.450.3289G}, edge-on Sbc galaxy \citep{1990ApJ...362..434C}. The source is also a member of the interacting group the Leo Triplet, together with the other two galaxies NGC~3623 and NGC~3627. \obj\ is a sub-LIRG (sub-luminous infrared galaxy, $10^{10}L_\odot<=L_\mathrm{IR}<10^{11}L_\odot$) with infrared luminosity $L_\mathrm{IR}=1.7\pm0.4\times10^{10}L_\odot$ \citep{2021A&A...649A.105F}. In particular, no evidence of an obscured AGN was found in \obj\ \citep{2021A&A...649A.105F}. The high infrared luminosity of \obj\ is interpreted as evidence of intense star formation. In addition, \obj\ satisfies the well-known FIR-radio correlation for star-forming galaxies, indicating that the central engine is dominated by star formation. \obj\ maintains wide-scale outflows, including a sub-kpc scale molecular gas outflow \citep{1996ApJ...464..738I, 2012ApJ...752...38T, 2014A&A...562A..21C, 2016MNRAS.463.2296R} and kpc-scale ionized gas outflows \citep{2010ApJ...711..818S, 2022A&A...660A.133H}. Again, X-ray observations with high enough resolution reveal a soft X-ray halo in the central region of \obj\ \citep{2002ApJ...568..689S, 2004ApJS..151..193S, 2012ApJ...752...38T}, which mimics a warm gas outflow. Through high-resolution radio observations \citep[][VLA A-array 15\,GHz and MERLIN 1.4\,GHz, respectively]{1990ApJ...362..434C, 1998MNRAS.300..656C}, a list of compact radio sources were identified as star-forming products. Therefore, combining radio and X-ray observations may help to confirm the nature of star forming galaxies \citep{2014ApJ...797...79W, 2015AJ....150...91P}. Unfortunately, previous radio observations were unable to resolve the central radio sources, so a higher resolution is necessary to determine their sizes and structures.

This work present VLBI 1.5\,GHz observational studies of the starburst galaxy \obj, and this observation was to search for the potential low luminosity AGN in \obj, probe the nuclear radio sources and constrain their sizes. The archival X-ray data of \obj\ are also analyzed to ensure a better understanding of the central engine of the galaxy \obj. In Section 2, we present observation and data reduction. We present the data reduction results in Section 3. In Section 4, we discuss the identification of the radio sources detected by VLBI, the model-fits of X-ray spectra, and the explanation of the X-ray and radio emissions. Finally, in Section 5, we give our conclusions. Throughout this work, we assume a distance of 10.6\,Mpc to \obj, which yields a size scale of 0.051\,pc\,mas$^{-1}$, and take the standard $\Lambda$CDM cosmology with $\Omega_\Lambda=0.71$, $\Omega_m=0.27$ and $T_{cmb}=2.732$\,K.

\begin{figure*}
\centering \includegraphics[scale=0.8]{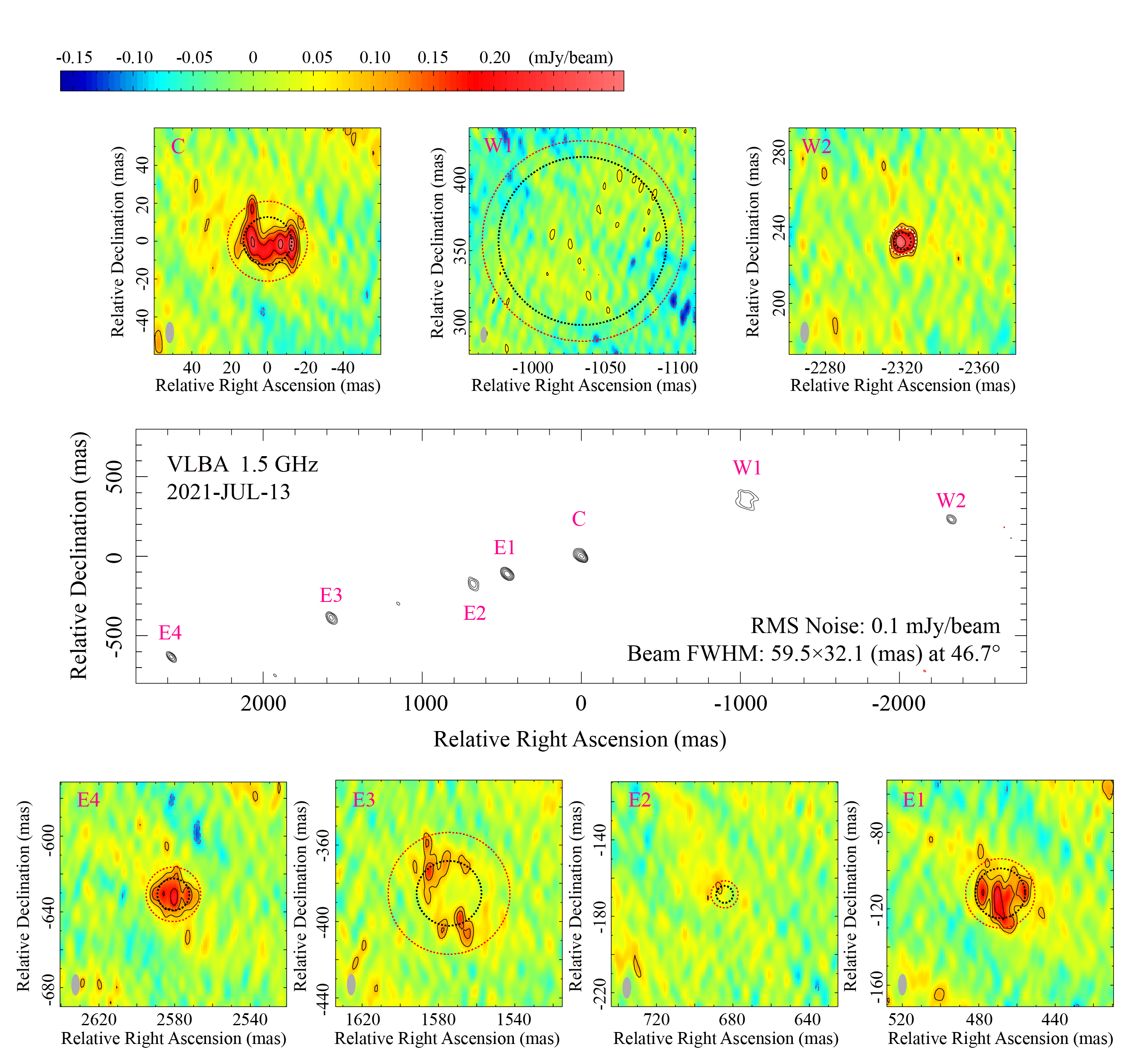}
\caption{VLBA 1.5\,GHz images of \obj. \textit{The middle panel:} the uv-tapered images of \obj; \textit{Upper and lower panels}: the full resolution images of each source presented in the middle panel, where the grey ellipses in the bottom left corner of each panel represent the full width at half-maximum (FWHM) of the restoring beam, they are all the same, i.e. $11.2\times4.47$ at $-0.63^\circ$. The contours are at 3$\sigma\times(-1, 1, 1.41, 2, 2.83,...)$, where $1\sigma$ noises are $0.1$ and $0.03\,\mathrm{mJy\,beam^{-1}}$ for the full resolution and uv-tapered contours, respectively. The black and red/white circles in each upper and lower panel indicate the component size obtained via Gaussian and spherical model-fits, respectively. \label{fig:vlbi}}
\end{figure*}

\begin{figure*}
\centering \includegraphics[scale=0.5]{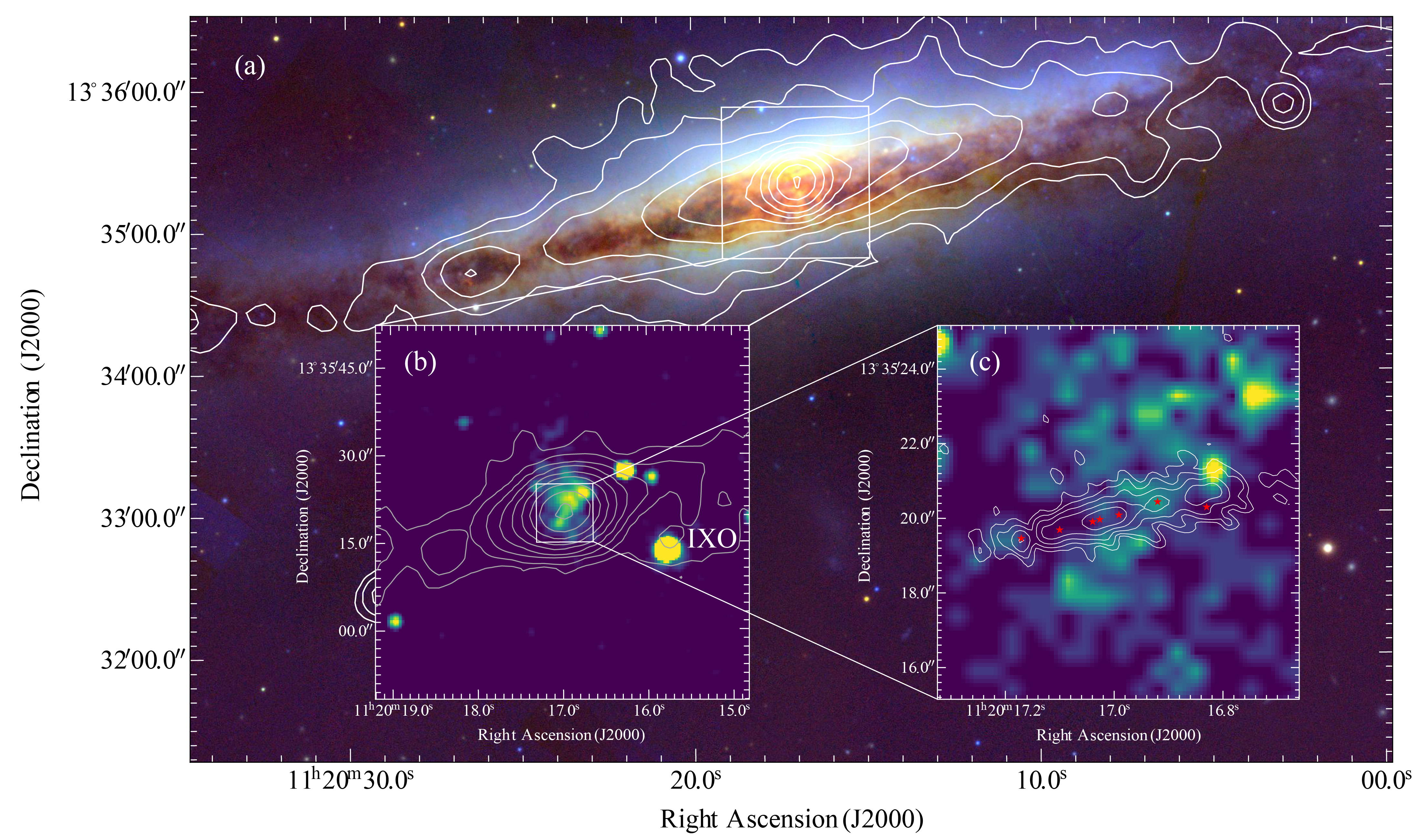}
\caption{A zooming-in to the central starburst region of \obj. (a) the $y$, $i$, $g$-band color-composite image from Pan-STARRS DR1 (PS1) is overlaid by the VLA D-array 1.5\,GHz contours; (b) the FIRST 1.4\,GHz contours overlay on the $0.5-7\,\mathrm{keV}$ $Chandra$ X-ray image and the bright point X-ray source is marked as IXO (intermediate-luminosity X-ray object) according to the identification by  \citet{2001ApJ...560..707S}; (c) VLA A-array 4.86\,GHz radio contours overlay on the $Chandra$ X-ray image, where red stars mark the positions of the seven sources detected by VLBA 1.5\,GHz (see contours in the middle panel of Figure \ref{fig:vlbi}). All the contours are at 3$\sigma\times(-1, 1, 2, 4, 8,...)$, where $1\sigma$ noises are $0.11$, $0.15$, $0.06$\,$\mathrm{mJy\,beam^{-1}}$, respectively. \label{fig:full}}
\end{figure*}

\section{Observation and data reduction}

\subsection{VLBA}

We observed \obj\ on 2021 July 13 with eight VLBA antennas at $L$-band ($1.545$\,GHz, hereafter using $1.5$\,GHz for short, the project code BA146). The total observation time is 2 hr with a data recording rate of 2\,Gbps. The observation was performed in phase-referencing mode, using J1118$+$1234 ($\alpha=\mathrm{11\mathrm{^h}18\mathrm{^m}57\mathrm{^s}.301434}$, $\delta=\mathrm{+12^\circ34^\prime41^{\prime\prime}.71798}$, J2000, i.e. $1.06^\circ$ to the target) as the phase reference calibrator. The VLBA data calibration was performed in the Astronomical Image Processing System (\softw{aips}), a software package that was developed by the National Radio Astronomy Observatory (NRAO) of USA \citep{2003ASSL..285..109G}, following the standard procedure. A priori amplitude calibration was performed using the system temperatures and antenna gain curves provided by each station. The Earth orientation parameters were obtained and calibrated using the measurements from the US Naval Observatory database, and the ionospheric dispersive delays were corrected from a map of the total electron content provided by the Crustal Dynamics Data Information System (CDDIS) of NASA \footnote{\url{https://cddis.nasa.gov}}. The opacity and parallactic angles were also corrected using the auxiliary files attached to the data. The delay in the visibility phase was solved using the phase reference calibrator J1118$+$1234. A global fringe-fitting on the phase-referencing calibrator was performed by taking the calibrator's model to solve miscellaneous phase delays of the target. The calibrated data of the target was then exported to \softw{difmap} \citep{1997ASPC..125...77S} for self-calibration and model-fit. Especially, we first performed a uv-tapper in \softw{difmap} by 0.1 at 5\,M$\lambda$ to construct a low-resolution image, where a two-dimensional Gaussian model-fit was used. Only sources with signal-to-noise ratios above $5$ were fitted. The middle panel of Figure \ref{fig:vlbi} shows the uv-tapered image of \obj. Next, the full-resolution image of each Gaussian component was established with \module{clean} in \softw{difmap}, and we especially consult the positions of components W1 and E2 from the Gaussian models. No self-calibration was applied to the target source due to the low signal-to-noise ratios. The full-resolution images for each source were created using natural weights; see subset images in Figure \ref{fig:vlbi}. Based on a well-established understanding of their nature, a model fit using an optically thin sphere was conducted to determine the sizes and flux densities of these radio sources (see Section \ref{sec:sigma-relation}). Furthermore, we obtain VLA B-array 1.4\,GHz image from the FIRST survey \citep[the Faint Images of the Radio Sky at Twenty-centimeters, ][]{1995ApJ...450..559B} \footnote{\url{http://sundog.stsci.edu/}} and VLA D-array 1.5\,GHz (observed on 1988-03-25, the project ID. AS300 and PI: G. Siemieniec) and A-array 4.86\,GHz \citep[observed on 1984-12-10, the project ID. AW122 and PI: A.E. Wehrle,][]{1987BAAS...19Q.718W} images from the NRAO Archive \footnote{\url{https://data.nrao.edu/}}, see the contours in Figure \ref{fig:full}.

We estimate uncertainties in coordinates by accounting for three main origins: (1) Positional uncertainties of phase-referencing calibrators. In phase-referencing observations, the coordinates of the target are referenced by a closed calibrator. The calibrator J1118$+$1234 was elected from the catalog rfc\_2022a in Astrogeo \footnote{\url{http://astrogeo.org/}} with precise position with accuracies $\Delta\alpha=0.10$\,mas in right ascension and $\Delta\delta=0.10$\,mas in declination; (2) Astrometric accuracy of phase-referenced observations. Primarily concerning the station coordinate, Earth orientation, and troposphere parameter uncertainties, which can be measured through the formula and parameters from \citet{2006A&A...452.1099P}. This portion contributes position errors: $\Delta\alpha\approx0.26$\,mas and $\Delta\delta\approx0.50$\,mas in this VLBA 1.5\,GHz observation; (3) Thermal error due to the random noise \citep[e.g.][]{2017AJ....153..105R}. This uncertainty can be characterized as $\sigma_t\sim\theta_\mathrm{B}/(2\times\sigma_{SNR})$, where $\theta_\mathrm{B}$ is the full-width at half maximum of the restoring beam and $\sigma_{SNR}$ is the signal-to-noise ratio. In this work, we measure the thermal error via a Gaussian model-fit in \softw{difmap}. In Table \ref{tab:radio}, we list the coordinates along with the estimated uncertainties of the coordinates for each VLBI source.

We estimate flux density uncertainties following the instructions described by \citet{1999ASPC..180..301F}. In this work, the integrated flux densities $S_i$ were extracted from the Gaussian model-fit in \softw{difmap} with the task \module{modelfit}, where a standard deviation in the model-fit was estimated for each source and considered as the fitting noise error. Additionally, we assign the standard 5\% errors originating from amplitude calibration of VLBA (see VLBA Observational Status Summary 2021A \footnote{\url{https://science.nrao.edu/facilities/vlba/docs/manuals/oss2021A}}).

\begin{figure}
\centering \includegraphics[scale=0.4]{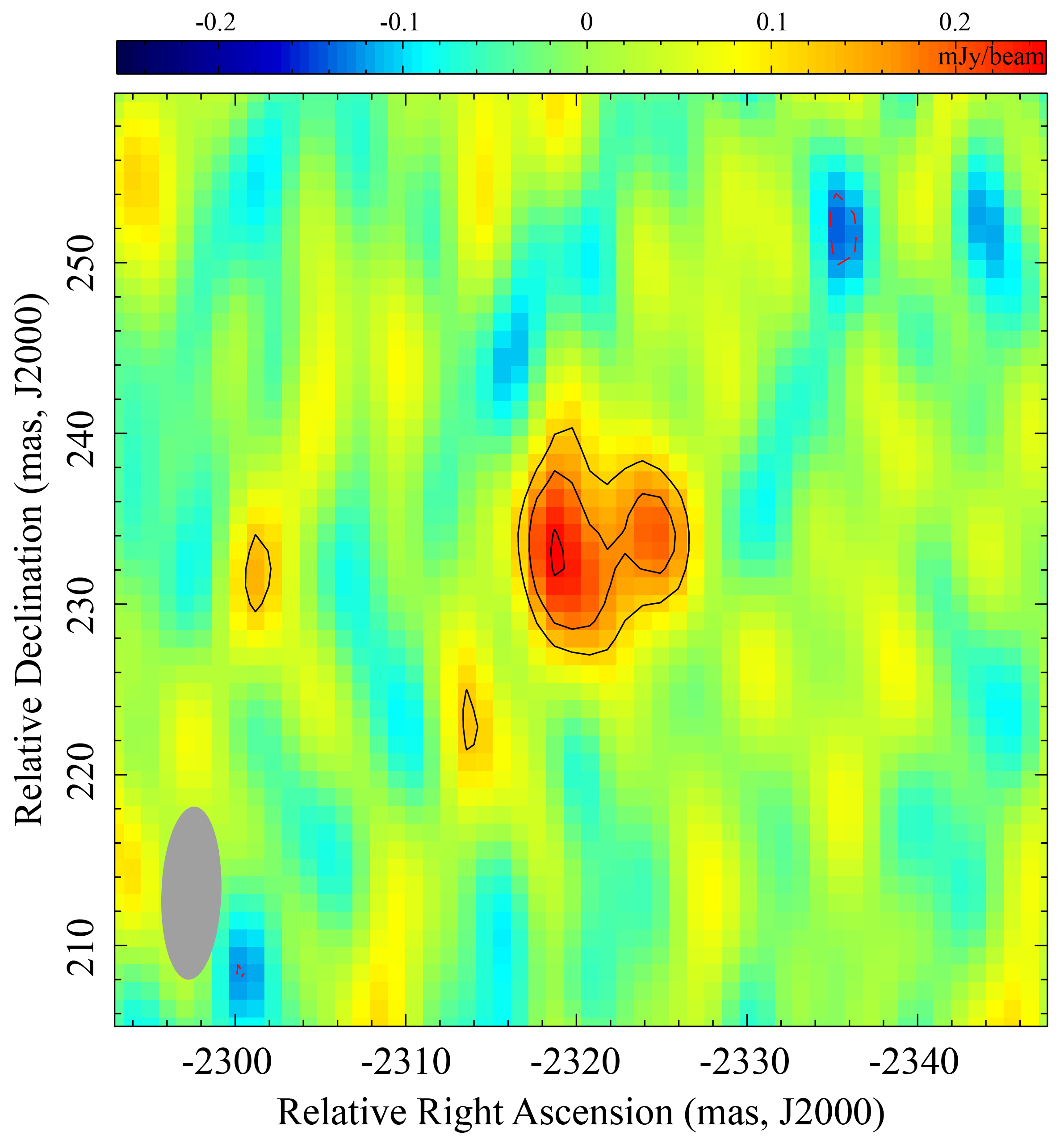}
\caption{The uniformly weighted VLBA 1.5\,GHz image of the source W2. The black solid contours represent positive values and the red dashed contours represent negative values. The contours are at 3$\sigma\times(-1, 1, 1.41, 2, 2.83,...)$, here $1\sigma$ noise is 0.04\,mJy\,beam$^{-1}$. The grey ellipse in the bottom left corner represents the full width at half-maximum (FWHM) of the restoring beam, it's $10.1\times3.51$\,mas at $-2.07^\circ$. \label{fig:w2}}
\end{figure}

\begin{deluxetable*}{ccccccccc}
\tablecaption{Radio observational properties of the seven sources in the nuclear region of \obj. \label{tab:radio}}
\tablehead{
\colhead{ID} &\colhead{$F_\mathrm{VLBA,1.5GHz}$} &\colhead{$d_\mathrm{VLBA,1.5GHz}$} &\colhead{$S_\mathrm{VLBA,1.5GHz}$} &\colhead{$\theta_\mathrm{VLBA,1.5GHz}$} &\colhead{$S_\mathrm{MERLIN,1.4GHz}$}&\colhead{$S_\mathrm{VLA-A,15GHz}$} &\colhead{$\alpha_\mathrm{1.4-15GHz}$} &\colhead{$\log{T_\mathrm{B}}$}\\
\colhead{} & \colhead{(mJy)}   & \colhead{(mas)} & \colhead{(mJy)} & \colhead{(mas)} & \colhead{(mJy)}     & \colhead{(mJy)}  & \colhead{} & \colhead{(K)}
}
\decimalcolnumbers
\startdata
C  &$2.41\pm0.22$ &$42.2\pm5.9$ &$2.67\pm0.21$&$25.4\pm1.2$&$5.7\pm2.2$&$2.4\pm0.2$ & $-0.37\pm0.21$ & $6.49\pm0.05$\\
E1 &$1.67\pm0.19$ &$36.5\pm6.7$ &$2.36\pm0.23$&$26.3\pm1.5$&$7.0\pm1.1$&$2.8\pm0.2$ & $-0.39\pm0.07$ & $6.41\pm0.06$\\
E2 &$0.78\pm0.12$ &$14.3\pm3.4$ &$0.81\pm0.12$&$8.9\pm0.8$&$4.9\pm1.1$&$2.3\pm0.1$ & $-0.32\pm0.10$ &$5.55\pm0.11$\\
E3 &$1.81\pm0.24$ &$64\pm14$    &$1.64\pm0.21$&$34.2\pm3.0$&$2.6\pm1.1$&$1.5\pm0.1$ & $-0.23\pm0.18$ &$6.02\pm0.09$\\
E4 &$1.19\pm0.16$ &$28.7\pm6.2$ &$1.21\pm0.15$&$16.8\pm1.4$&$1.3\pm1.0$&$0.4\pm0.1$ & $-0.50\pm0.34$ &$6.51\pm0.09$\\
W1 &$2.18\pm0.95$ &$139\pm84$   &$4.83\pm3.70$&$117\pm33$&$2.3\pm0.6$&$0.8\pm0.1$ & $-0.45\pm0.13$ &$5.42\pm0.41$\\
W2 &$0.79\pm0.12$ &$14.2\pm3.4$ &$0.81\pm0.12$&$8.9\pm0.8$&$<0.4$&$<0.21$ & $-0.5^\dagger$ &$6.88\pm0.10$\\
\enddata

\tablecomments{Columns give (1) Source ID, (2-3) VLBA 1.5\,GHz flux density and diameter with the optically thin sphere model-fit, (4-5) VLBA 1.5\,GHz flux density and FWHM of Gaussian model-fit, (6-7) MERLIN 1.4\,GHz and VLA 15\,GHz flux density, (8) 1.4 (MERLIN) - 15\,GHz (VLA) spectral index, and (9) VLBA 1.5\,GHz brightness temperature estimated using the parameters from Gaussian model-fit.}
\tablecomments{$\dagger$: The source was not detected by both MERLIN 1.4\,GHz and VLA A-array 15\,GHz observations, here we assume a general spectral index for SNRs.}

\end{deluxetable*}

\begin{deluxetable*}{ccccccc}
\tablenum{2}
\tabletypesize{\normalsize}
\tablecaption{Summary of the radio observations of the target \obj. \label{tab:obs}}
\tablewidth{0pt}
\tablehead{
\colhead{Telescope} & \colhead{Epoch} & \colhead{Frequency} & \colhead{Bandwidth} & \colhead{$\theta_\mathrm{maj}$} & \colhead{$\theta_\mathrm{min}$} & \colhead{PA}\\
\colhead{}          & \colhead{}      & \colhead{(GHz)}     & \colhead{(MHz)}     & \colhead{(mas)}          & \colhead{(mas)}          & \colhead{($\circ$)}
}
\decimalcolnumbers
\startdata
MERLIN         & 1996-1   & 1.4 &$ 16    $&$ 340    $&$300     $&$7.0$  \\
MERLIN$^\star$ & 1996-1   & 1.4 &$ 16    $&$ 260    $&$160     $&$7.0$  \\
VLA-A          & 1986-5   & 15  &$ 100   $&$ 150    $&$140     $&$-12$        \\
VLBA$^\bullet$ & 2021-7-2   & 1.5 &$ 256   $&$ 59.5   $&$32.1    $&$46.7$       \\
VLBA           & 2021-7-2   & 1.5 &$ 256   $&$ 11.2   $&$4.47    $&$-0.63$      \\
\enddata
\tablecomments{Columns give (1) telescope, 
(2) observing epoch, (3) frequency, (4) bandwidth, and (5$-$7) major, minor, and position angle of the beam. \\ Parameters are estimated under natural weights except for \\ $\star$: the uniform weight. \\ $\bullet$: the uv-taper.}
\end{deluxetable*}

\subsection{X-ray}
\obj\ was observed by \telescope{Chandra} ACIS-S on December 2, 2000 \citep[ObsID: 2039, PI: David K. Strickland;][]{2001ApJ...560..707S} with FAINT mode and a total exposure time of 57.96\,ks. We retrieved the data from the $Chandra$ data archive, performed data reduction, and constructed the X-ray spectrum with the Chandra Interactive Analysis of Observations software package \citep[\softw{ciao},][]{2006SPIE.6270E..1VF} \footnote{\url{https://cxc.cfa.harvard.edu/ciao/}} version 4.13. The data was initially reprocessed according to the Chandra Calibration Database (\softw{caldb}) \footnote{\url{https://cxc.cfa.harvard.edu/caldb/}} version 4.9.6 standards using \module{chandra$\_$repro}. We extracted the X-ray spectrum using the task \module{specextract}. Previous X-ray observational attention of \obj\ is an off-nuclear ($20^{\prime\prime}$ from the galaxy center) point X-ray source \citep[see][]{2001ApJ...560..707S}, other than the diffuse X-ray emissions in the central region of this galaxy. This point X-ray source is bright ($L_\mathrm{X,0.3,8keV}=1.1\times10^{40}\,\mathrm{erg\,s^{-1}}$) and strongly variable, thus belongs to ultraluminous X-ray sources \citep{2023NewAR..9601672K} \citep[but it was initially called intermediate-luminosity X-ray object, IXO, by][]{2001ApJ...560..707S}. In this work, we only concern the nuclear X-ray emissions of this galaxy and we use \module{specextract} to extract the spectrum of \obj\ from a $6^{\prime\prime}$-radius aperture centered on the nuclear X-ray emissions; the background spectrum is extracted from a nearby source-free region. We used photons with energy ranging from 0.5 to 7.0\,keV and grouped the spectrum into 20 photons in each bin.

For cross-checking purposes, we also retrieved the data from \telescope{XMM-Newton}. The data was obtained on November 27, 2000 (ObsID: 0110980101, PI: Fred Jansen) with 65\,ks. We used the data from all three X-ray detectors: MOS-1, MOS-2, and PN equipped on the European Photon Imaging Cameras (EPIC), and the clean time intervals are 42.2, 42.3, and 32.2\,ks, respectively. The EPIC data processing was done with the XMM-Newton Science Analysis System (SAS) \footnote{\url{https://www.cosmos.esa.int/web/xmm-newton/sas}} 16.0.0. The tasks \module{emproc} and \module{epproc} were used to reprocess the data. To extract the spectrum, we used the same region as with \telescope{Chandra} and the energy range of 0.1 and 10 keV. We used the tasks \module{rmfgen} and \module{arfgen} to create a response matrix file (RMF) and auxiliary response file (ARF).

\section{Results}
In Figure \ref{fig:vlbi}, we show VLBA 1.5\,GHz images in both high/full (subset images in the top and bottom panels) and low resolution (the middle panel). Based on the VLBA 1.5\,GHz observation we identify seven point-like sources under a restoring beam of $59.5\times32.1$\,mas, while two of them (W1 and E2) are too weak and diffuse to be detected in the full-resolution image, and the beam size of the full-resolution image is $11.2\times4.47$\,mas correspondingly. A two-dimensional Gaussian model-fit is employed to obtain the integrated flux density and the sizes (full width at half maximum of Gaussian models, FWHM) of the compact sources; the model-fitting results are listed in Table \ref{tab:radio}.

Taking the parameters estimated via Gaussian model-fit for each VLBA source, we can estimate the brightness temperatures using the formula \citep[e.g.][]{2005ApJ...621..123U} 
\begin{equation}\label{eq:bt}
T_\mathrm{B}=1.8\times10^9(1+z)\frac{S_i}{\nu^2\theta^2}~\mathrm{(K)},
\end{equation}
where \(S_i\) is the integrated flux density of each Gaussian model component in mJy (column 2 of Table \ref{tab:radio}), $\theta$ is $\mathrm{FWHM}$ of the Gaussian model component in mas (column 6 of Table \ref{tab:radio}), \(\nu\) is the observing frequency in GHz, and \(z\) is the redshift. The estimated 1.5\,GHz brightness temperatures are listed in column 7 of Table \ref{tab:radio}. Here, the estimated size $\theta$ is reliable by comparing it with the full-resolution image of each source. Therefore, it ensues a robust estimation of the radio brightness temperatures. Sources E2 and W1 have the brightness temperatures $\sim10^5$\,K, while the others all have brightness temperatures $>10^6$\,K including the highest one W2 (reaching $10^{6.88}$\,K). 

We cross-match VLBA and MERLIN sources within 1 beam size ($0.26\times0.16$ at $7.0^\circ$) of the MERLIN 1.4\,GHz observation \citep{1998MNRAS.300..656C}. Six out of the seven VLBA sources (i.e. excepting source W2) have MERLIN counterparts, and VLA A-array 15\,GHz sources as well \citep{1990ApJ...362..434C, 1998MNRAS.300..656C}, see Table \ref{tab:radio}. We list the spectral index estimated between MERLIN 1.4\,GHz and VLA A-array 15\,GHz by \citet{1998MNRAS.300..656C} in column 8 of Table \ref{tab:radio}. W2 is either a new or a compact (but weak, therefore it is submerged by diffuse foreground radio emissions in previous VLA 15\,GHz and MERLIN 1.4\,GHz observations) radio source. In Figure \ref{fig:full}, we present the optical image of \obj\ with radio contours overlaid and the distribution of X-ray emission.

\begin{deluxetable*}{llcccccccc}
\tablenum{3}
\tabletypesize{\scriptsize}
\tablecaption{Model-fitting results of the X-ray spectra.\label{tab:xfit}}
\tablewidth{0pt}
\tablehead{
\colhead{Telescope} & \colhead{XSPEC Model} & \colhead{$N_{\rm H}$} & \colhead{$\Gamma$} & \colhead{PL Norm} & \colhead{$kT$} & \colhead{Mg} & \colhead{Si} & \colhead{$\mathrm{Norm_2}$} & \colhead{$\chi^2$/d.o.f.} \\ 
\colhead{} & \colhead{}  & \colhead{$(10^{22}\,\rm cm^{-2})$}  & \colhead{} & \colhead{$(10^{-5})$} & \colhead{(keV)} & \colhead{$(Z_\odot)$} & \colhead{$(Z_\odot)$} & \colhead{$(10^{-4})$} & \colhead{}
}
\decimalcolnumbers
\startdata
\telescope{Chandra}     & $\mathrm{wabs*powerlaw}$           & $0.40\pm0.07$ & $1.73\pm0.18$ & $1.46\pm0.27$ &               &               &               &               & $39.68/38$ \\
\telescope{Chandra}     & $\mathrm{wabs*(powerlaw+vmekal)}$  & $0.71\pm0.73$ & $1.61\pm0.38$ & $1.28\pm0.73$ & $0.27\pm0.33$ & $9.49\pm13.2$ & $51.1\pm76.8$ & $\sim1.71$   & $25.0/38$ \\
\telescope{Chandra}     & $\mathrm{wabs*(powerlaw+vapec)}$   & $0.81\pm0.45$ & $1.64\pm0.28$ & $1.36\pm0.55$ & $0.27\pm0.15$ & $5.35\pm4.18$ & $29.9\pm31.2$ & $\sim1.52$   & $23.21/38$ \\
\telescope{XMM-Newton}         & $\mathrm{wabs*powerlaw}$           & $0.63\pm0.13$ & $2.10\pm0.21$ & $4.31\pm1.05$ &               &               &               &               & $16.91/30$ \\
\telescope{XMM-Newton}         & $\mathrm{wabs*(powerlaw+vmekal)}$  & $1.26\pm0.41$ & $2.34\pm0.25$ & $6.19\pm2.04$ & $0.45\pm0.18$ & $\sim0.79$        & $2.03\pm3.43$ & $1.33\pm1.94$ & $9.56/30$ \\
\telescope{XMM-Newton}         & $\mathrm{wabs*(powerlaw+vapec)}$   & $1.34\pm0.43$ & $2.27\pm0.25$ & $5.69\pm1.94$ & $0.39\pm0.29$ & $\sim0.14$        & $1.77\pm2.61$ & $2.32\pm5.08$ & $9.56/30$ \\
\enddata

\tablecomments{Columns give (1) telescope, (2) \softw{xspec} models for fit, (3) column density of the neutral Hydrogen, (4$-$5) photon index and normalization of the power-law model, (6$-$9) temperature, abundances of the elements Mg and Si, and the normalization of \module{vmekal} or \module{vapec} model, and (10) fit $\chi^{2}$ / degrees of freedom.}
\end{deluxetable*}

\section{Discussion}

\subsection{Properties of the nuclear radio sources\label{subsec:nuclear}}
All the seven radio sources in \obj\ have radio brightness temperatures $>10^5$\,K, which are thought to have non-thermal origins \citep{1992ARA&A..30..575C}. \citet{1998MNRAS.300..656C} suggest that six out of the seven radio sources, which have both MERLIN 1.4\,GHz and VLA A-array 15\,GHz counterparts, are all SNRs based on the spectral indices of $-0.23$ to $-0.50$ between 1.4 and 15\,GHz. The structures of sources C and E3 may be consistent with partial shells in the full-resolution VLBA images, which further suggests that both of them are SNRs. W2 is the most compact one in our VLBA observation with the highest brightness temperature $\log{(T_b/\mathrm{K})}=6.88$ and the compact structure in the full resolution VLBA image. In addition, we produce a uniformly weighted image of the source W2 (see Figure \ref{fig:w2}), and it resembles a partial shell of radio SN with asymmetric structure \citep[e.g.][]{1997ApJ...479..845G, 2005ApJ...628L.131M, 2002PASA...19..207M}. Six out of the seven VLBA sources (except W2) distribute along the galaxy disk. Depending on an exponential initial mass function and the fact that most massive stars will become SNe, the linear distribution of the SNe in \obj\ hints at an extremely thin, edge-on, and dense inner ($\sim250$\,parsec) star-forming disk. Furthermore, the warping of the galaxy's outer disk is one plausible explanation for the positional discrepancy between W2 and the plane defined by the other six sources and the galaxy.

Based on the transient nature and the partially shell-like morphology, W2 is possibly a radio SN. The lack of an X-ray point source visible in the \telescope{Chandra} image makes it less likely that the object is an X-ray binary or an AGN. The non-detection of W2 in both VLA and MERLIN observations in 1986 and 1996, respectively, indicates that either W2 had low radio flux densities at these epochs or it just exploded after 1996. By integrating the Salpeter initial mass function with a power-law slope of $\alpha\approx-2.35$ \citep[see][]{1955ApJ...121..161S,2001MNRAS.322..231K,2003PASP..115..763C,2011MNRAS.415.1647G} and assuming that stars with masses greater than $8\,M_\odot$ will eventually become SNe \citep{2003ApJ...591..288H}, the SN rate can be estimated. This calculation results in a 0.2\% SN rate triggered by the star formation rate. The SFR of \obj\ is $3.2\,M_\odot\,\mathrm{yr^{-1}}$ \citep{2012ApJ...752...38T} or $4.8\,M_\odot\,\mathrm{yr^{-1}}$ \citep{2015ApJ...815..133S}, which yields a SN rate of 0.0064 or 0.0096 year$^{-1}$, respectively. Although these estimates are only approximate, as only a small portion of SNe are radioactive and above the VLBA 1.5 GHz detection limit, it is consistent with the fact that W2 is a nascent SN that has exploded between 2021 (the epoch of VLBA 1.5 GHz observation) and at least 1986.


\begin{figure*}
\centering
\includegraphics[width=0.9\textwidth]{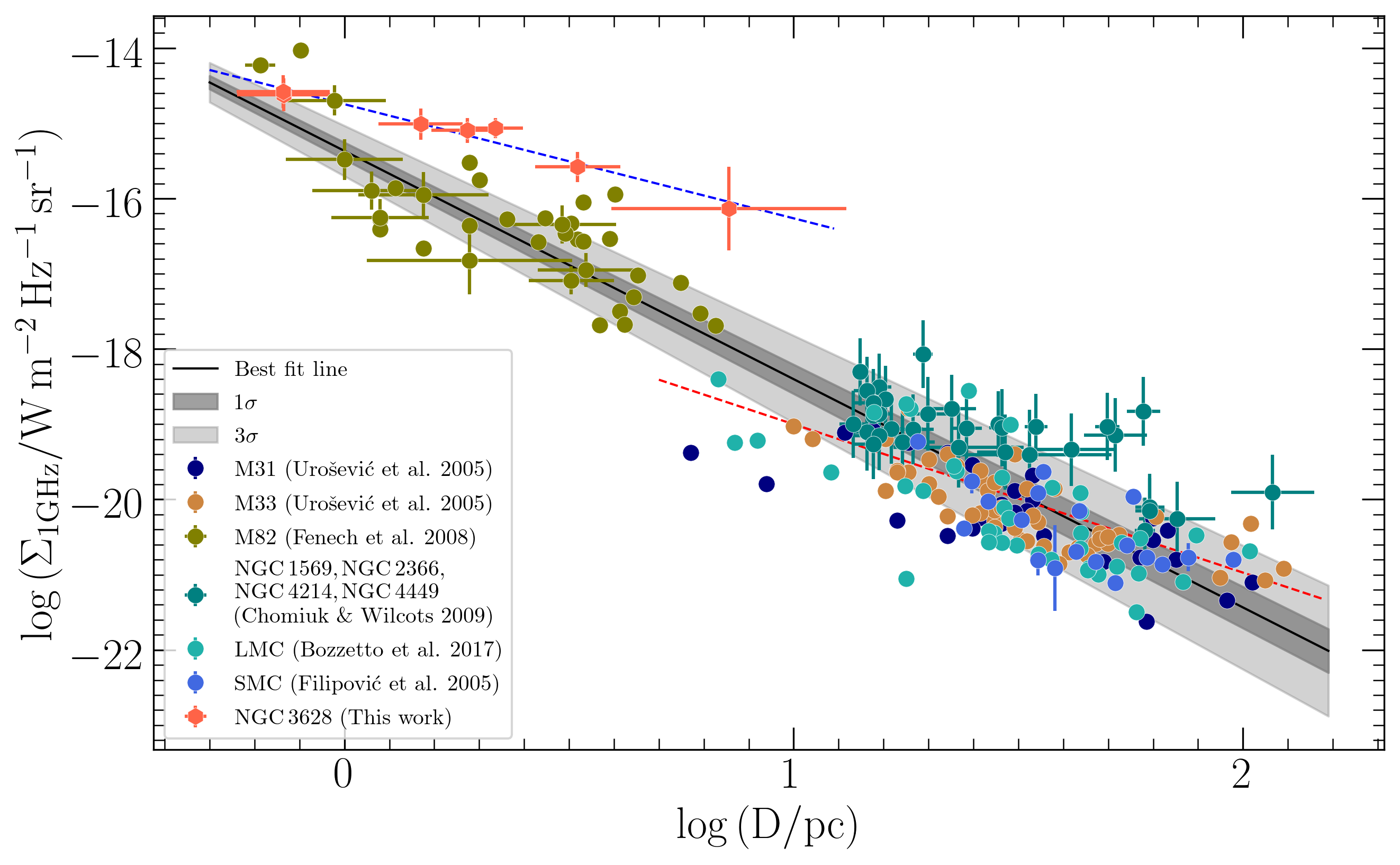}
\caption{The 1\,GHz surface brightness versus diameter relation for SNRs. The solid black line represents the best fit of the full SNR samples and the grey belts are $1\sigma$ and $3\sigma$ intervals. The blue- and red-dashed lines are the best fits for the SNR sample in \obj\ and in local galaxies except for M82 and \obj, respectively. \label{fig:sd}}
\end{figure*}

\subsection{The $\Sigma-D$ relation for SNRs} \label{sec:sigma-relation}
The strength of the synchrotron emissions of the expanding SNRs follows the relation between radio surface brightness $\Sigma$ and diameter $D$ (the $\Sigma-D$ relation). This relation was initially derived by \citet{1960SvA.....4..243S} via a theoretical deduction and has the form 
\begin{equation}
    \Sigma_\nu=AD^{-\beta},
\end{equation}
where parameter $A$ depends on the properties of the SN explosion and surrounding interstellar medium (ISM): the SN energy of the explosion, the mass of ejecta, the density of the ISM, and the strength of the magnetic field, etc., while the slope $\beta$ is thought to be independent of these properties \citep{2005MNRAS.360...76A}. Subsequent theoretical studies \citep[e.g.][]{2004A&A...427..525B, 2005MNRAS.360...76A, 2013ApJS..204....4P} show that the $\Sigma-D$ relation is primarily applicable to shell-structure SNRs and obtain the slope $\beta$ in the range of 2$-$5.75. Generally, parameters $A$ and $\beta$ are observationally constrained by fitting the data to a sample of SNRs of known distances \citep[e.g.][]{2005A&A...435..437U, 2013ApJS..204....4P, 2017ApJS..230....2B}. Establishing a universal $\Sigma-D$ correlation provide a way to study the properties of SNR population in specific galaxies and interstellar environment. Furthermore, \citet{1960SvA.....4..243S} also proposed to use this relation as a method for determining SNR distances.

In order to constrain the parameters in $\Sigma-D$ relation, observations should have a sufficient resolution to measure the diameters of SNRs \citep[they usually maintain a size of sub-parsec to parsec scale, e.g.,][]{2008MNRAS.391.1384F, 2009AJ....137..537L, 2011ApJ...740...95B, 2012A&A...539A.134B, 2014AJ....147....5R} in star-forming galaxies. Based on the SN/SNR nature of the seven sources in \obj, we fit the optically thin sphere model to the uv-data for each source to obtain integrated flux density and size. Firstly, we perform the model fit on the full-resolution image to get the models of sources C, E1, E3, E4, and W2; then model the sources W1 and E2 on the uv-tapered data. The results are listed in columns 2 and 3 of Table \ref{tab:radio}. It can be found that the flux densities from the optically thin sphere model-fit (column 2 of Table \ref{tab:radio}) are consistent with those from the Gaussian model-fit (column 4 of Table \ref{tab:radio}), while the diameters of sphere models (column 3 of Table \ref{tab:radio}) are systematically larger than the FWHM of Gaussian models (column 5 of Table \ref{tab:radio}). This can be identified by marking these two sizes on the full-resolution images of each component (top and bottom panels of Figure \ref{fig:vlbi}). Obviously, the diameter of the sphere model indicates the outer boundary, while the FWHM of the Gaussian model indicates the size of the brightest emitting features. The optically thin sphere model is retrieved based on the identification of these sources as SNRs, and therefore it is more applicable to representing the source sizes.

We can fit the empirical $\Sigma-D$ correlation by establishing a well-resolvable SNR sample. In this work, we involve SNRs that are detected in several nearby galaxies \citep{2005A&A...435..437U, 2005MNRAS.364..217F, 2008MNRAS.391.1384F, 2009AJ....137.3869C, 2017ApJS..230....2B}, however, we decide to not use the SNR sample from Arp220 due to the poor resolutions \citep{2011ApJ...740...95B}. For SNRs in nearby low-redshift galaxies, we estimate 1\,GHz surface brightness $\Sigma_\mathrm{1GHz}\,(\mathrm{W\,m^{-2}\,Hz^{-1}\,sr^{-1}})$ through the formula
\begin{equation}
\Sigma_\mathrm{1GHz} = 5.417\times10^{-16}\frac{S_\mathrm{1GHz}}{\theta^2(1+z)^{(1+\alpha)}}, 
\end{equation}
where $S_\mathrm{1GHz}$ is the integrated 1\,GHz flux density in Jansky (Jy), $\theta$ is the diameter in arcsecond, $\alpha$ is the spectral index. The radio surface brightness doesn't depend on the distance to the SNRs. Our VLBA data has been extrapolated to 1\,GHz by using the spectral index in column 8 of Table \ref{tab:radio}.

Figure \ref{fig:sd} shows the surface brightness distribution along the diameters of SNRs, obviously, the SNRs which are detected in \obj\ follow the trend of the full sample. We fit a power law to the full SNR samples and get the correlation 
\begin{equation}
    \Sigma_\mathrm{1GHz}=(4.23\pm1.07)\times10^{-16}D^{-3.03\pm0.08}. 
\end{equation}
This fit shows a good regression and obtains the correlation slope $\beta=3.03\pm0.08$, which is slightly flatter than the fit by \citet{2017ApJS..230....2B}, i.e. $\beta=3.60\pm0.15$. This slight difference is due to the use of the new SNR sample of \obj\ in this work and the removal of the SNRs in Arp220 for fit. Fitting the SNR sample in \obj\ gives a power-law slope $\beta=1.51\pm0.12$. This slope is comparable with the value $\beta=1.97\pm0.16$ obtained for SNRs in the local galaxy samples except for M82 and \obj. The slope for SNRs in \obj\ is consistent with that in M31 and M33 \citep[$\beta=1.67$ and $\beta=1.77$, respectively][]{2005A&A...435..437U}, and NGC~1569 \citep[$\beta=1.26\pm0.2$,][]{2022MNRAS.513.1755S}. It should be noted that the slope $\beta\approx2$ is generally a statistical rather than a physical cause \citep{2005MNRAS.360...76A, 2022MNRAS.513.1755S}. However, the positive correlation between the surface brightness of SNRs and the ISM density, and the type of SNRs would naturally result in an individual $\Sigma-D$ relation for each extragalactic SNR sample. Furthermore, a high-resolution radio observation is important to measure the sizes of each SNR and constrain the $\Sigma-D$ relation.

\subsection{Origin of the nuclear diffuse X-ray emission}

\begin{figure}
\centering
\includegraphics[width=0.48\textwidth]{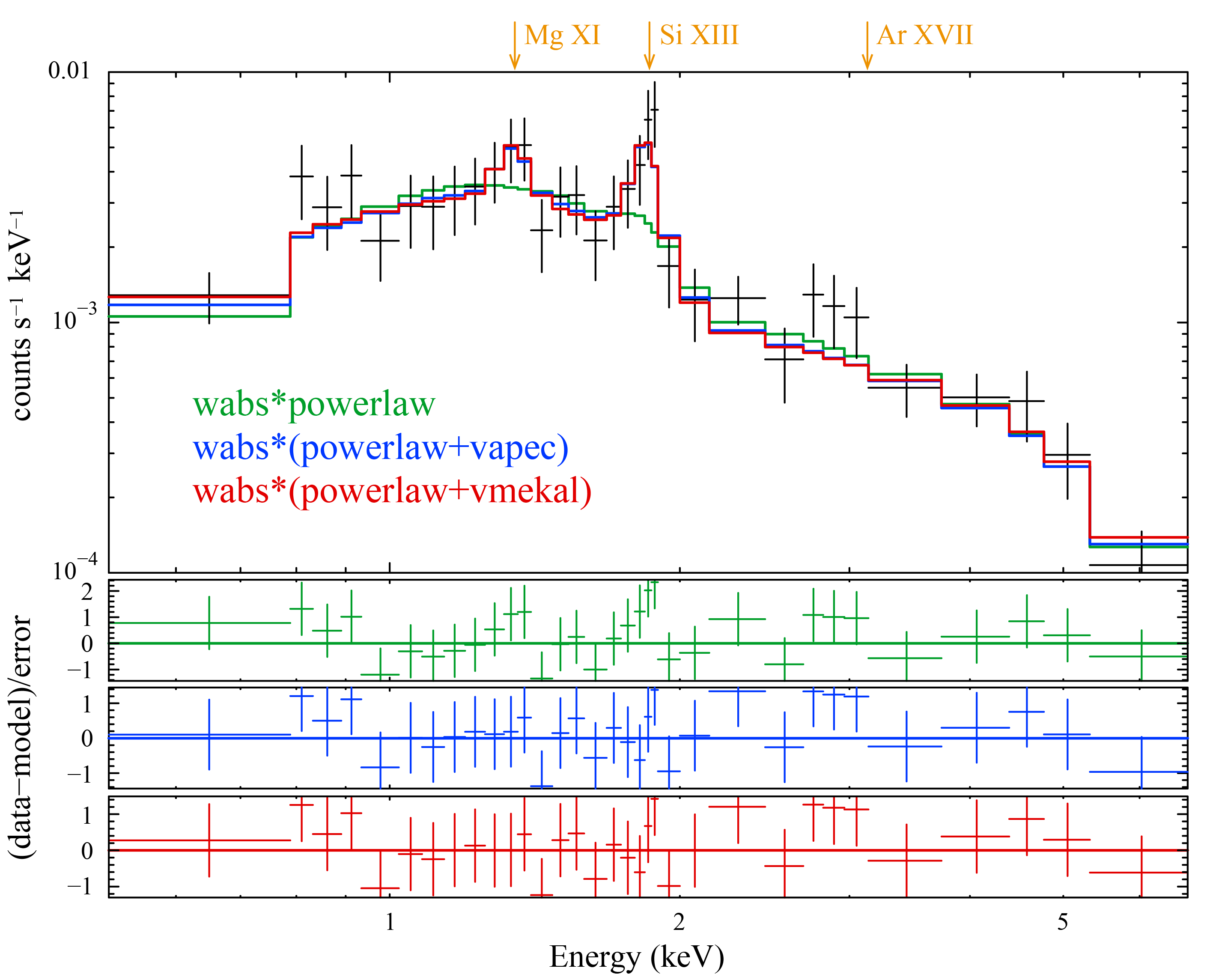}
\caption{\telescope{Chandra} spectrum and model-fits. The solid lines show the best-fitting models to the data in the energy range 0.5-7\,$\mathrm{keV}$, and each lower panel shows the corresponding residuals. Models from the top to the bottom panels are \texttt{wabs}$*$\texttt{powerlaw}, \texttt{wabs}$*$\texttt{(powerlaw+vapec)}, \texttt{wabs}$*$\texttt{(powerlaw+vmekal)}, respectively. \label{fig:chandra}}
\end{figure}

We extract \telescope{Chandra} X-ray spectrum directly from the central $12^{\prime\prime}$ region, which includes both diffuse and point-like X-ray sources (see Figure \ref{fig:full}). In starburst galaxies, point X-ray emission may originate from X-ray binaries \citep[e.g.][]{2014ApJS..212...21L} or AGNs \citep[e.g.][]{2003ApJ...582L..15K}. Therefore, we firstly fitted an absorbed power law model to the Chandra spectrum (the green line in Figure \ref{fig:chandra} and the parameters are listed in Table \ref{tab:xfit}), which follows the overall trend of the data in the count rate versus energy plot. Interestingly, by subtracting the continuum emission (the power law), we see several narrow features which overlap with the emission lines Mg\,\textsc{xi} and Si\,\textsc{xiii} (the rest-frame energies are 1.331 and 1.839\,keV, respectively), and possibly Ar\,\textsc{xvii} (3.104\,keV) and a combination of O \textsc{vii} (spans 0.653 to 0.654\,keV), Fe\,\textsc{xvii} (spans 0.725 to 0.739\,keV) and Ne\,\textsc{ix} (spans 0.905 to 0.922\,keV) lines. These lines are abundant in SNRs and are clear features of their X-ray spectra \citep[e.g.][]{1994PASJ...46L.151H, 2000ApJ...528L.109H, 2002ApJ...574..166M, 2006ApJ...645.1373B, 2010PNAS..107.7141B, 2020PASJ...72...65S}.

Therefore, we added a thermal plasma component to model the \telescope{Chandra} and \telescope{XMM-Newton} spectra. Two plasma models are individually added to the spectral fitting: collisional ionization equilibrium gas (\module{apec} model in \softw{xspec}) and hot diffuse MeKal gas (\module{vmekal} model in \softw{xspec}). In the spectral fitting, we only fit the abundances of the elements O, Fe, Ne, Mg, Si, and Ar, and fix the abundances of the other elements. However, we can only constrain the abundances (in units of solar abundance $Z_\odot$) of the elements Mg and Si (see Table \ref{tab:xfit}), which indicate a supersolar abundance. These two models well represent the X-ray spectra along with the line features, which give better $\chi^2_\mathrm{red}$ of 0.65 and 0.61 for models \module{wabs*(powerlaw+vmekal)} and \module{wabs*(powerlaw+vapec)} for the \telescope{Chandra} spectrum, respectively. The over-fitting indicated by the $\chi^2_\mathrm{red}$ is likely due to the limited number of data points.

Here the power-law component accounts for the possible contributions from XRBs and AGNs. Given that the radio emission of \obj\ is dominated by star-forming activity, we can explore the correlation between X-ray and radio luminosity for \obj. This correlation is regulated by the fact that both the radio and X-ray emissions are from star-forming activities, especially the X-ray emissions, are dominated by XRBs. We use a comparison sample of 43 star-forming galaxies with detection in both radio and X-ray bands from \citet{2012MNRAS.420.2190V}. A strong correlation was found in the comparison sample between the hard X-ray luminosity $\log{L_\mathrm{X,2-10keV}}$ in $\mathrm{erg\,s^{-1}}$ and 1.4\,GHz flux density $\log{L_\mathrm{1.4GHz}}$ in $\mathrm{W\,Hz^{-1}}$.
\begin{equation}
    \log{L_\mathrm{X,2-10keV}}=(1.04\pm0.05)\log{L_\mathrm{1.4GHz}}+(17.68\pm1.15), 
\end{equation}
and the slope is very close to unity for local star-forming galaxies. Furthermore, we fit the correlation between the soft X-ray luminosity $\log{L_\mathrm{X,0.5-2keV}}$ in $\mathrm{W\,Hz^{-1}}$ and 1.4\,GHz flux density for the comparison sample and find a correlation 
\begin{equation}
    \log{L_\mathrm{X,0.5-2keV}}=(0.89\pm0.06)\log{L_\mathrm{1.4GHz}}+(20.71\pm1.47). 
\end{equation}
We use the correlations here to estimate soft and hard X-ray luminosities for \obj\ and we take the FIRST 1.4\,GHz luminosity $\log{(L_\mathrm{1.4GHz}/\mathrm{W\,Hz^{-1}})}=21.58$, where the emission region constrained through FIRST observation well overlaps with that through $Chandra$ (see panel b of Figure \ref{fig:full}). The resulting $\log{(L_\mathrm{X,0.5-2keV}/\mathrm{erg\,s^{-1}})}=39.91\pm1.95$ and $\log{(L_\mathrm{X,2-10keV}/\mathrm{erg\,s^{-1}})}=40.12\pm1.57$, which are $\sim1.32$ and $\sim1.08$-dex higher than the intrinsic soft $\log{(L_\mathrm{X,0.5-2keV}/\mathrm{erg\,s^{-1}})}=38.58\pm0.01$ and hard X-ray luminosity $\log{(L_\mathrm{X,2-10keV}/\mathrm{erg\,s^{-1}})}=39.03\pm0.05$ of \obj, respectively, though with high uncertainties.

\begin{figure}
\centering 
\includegraphics[width=0.48\textwidth]{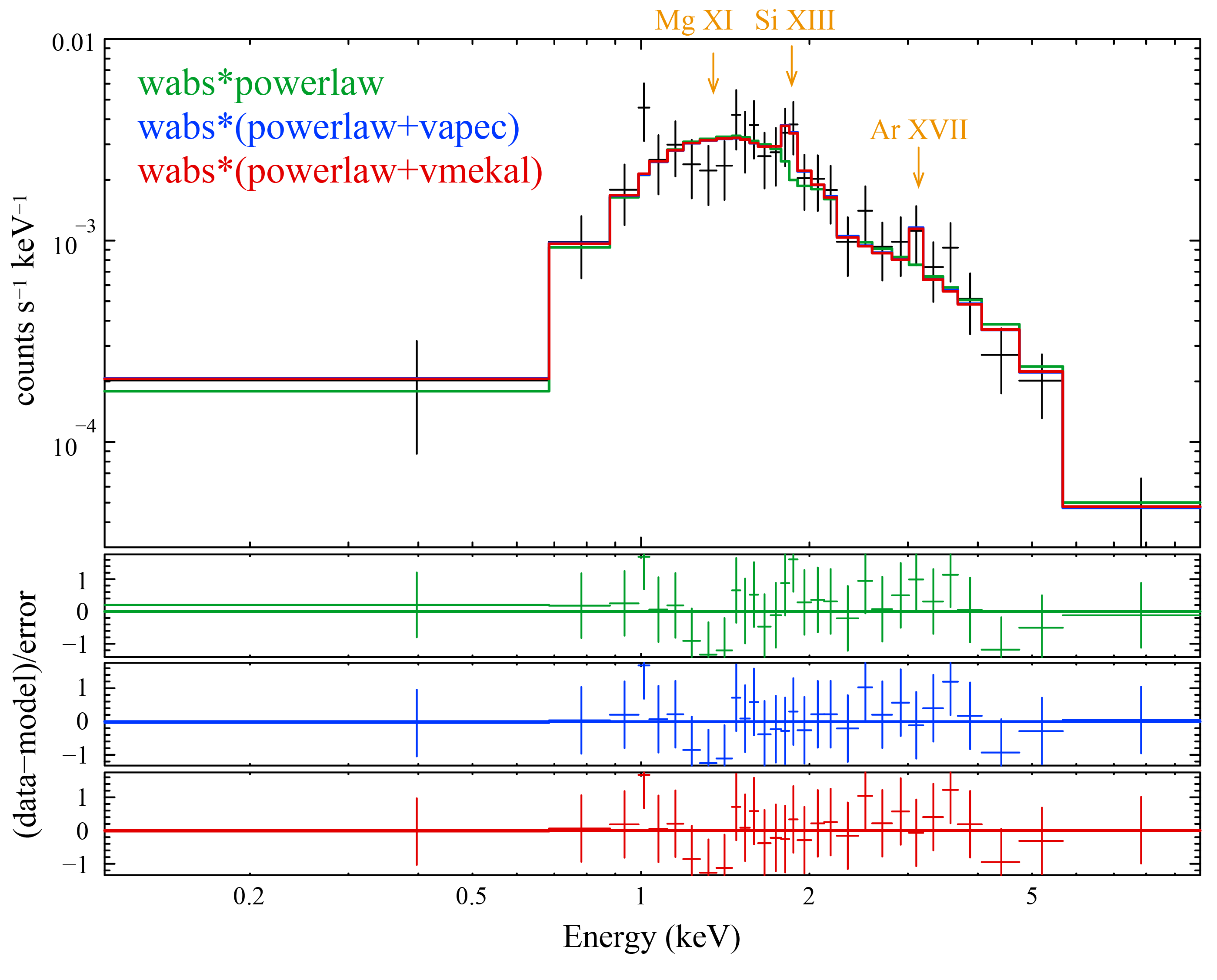}
\caption{\telescope{XMM-Newton} $0.1-10$\,keV spectrum of \obj\ and model-fits. The green, blue, and red lines represent the best fits with the model \texttt{wabs}$*$\texttt{powerlaw}, \texttt{wabs}$*$\texttt{(powerlaw+vapec)}, \texttt{wabs}$*$\texttt{(powerlaw+vmekal)}, respectively. The corresponding residuals of the best fits are shown in the bottom panel. \label{fig:xmm}}
\end{figure}

\tabletypesize{\small}
\begin{deluxetable}{ccc}
\tablenum{4}
\tablecaption{Model-fitting parameters for \telescope{Chandra} $0.5-7$\,keV spectrum with \softw{xspec} module \texttt{wabs}$*$\texttt{(vrnei+vmekal)}. \label{tab:fit_vv}}
\tablehead{\colhead{Model component} & \colhead{Paramerer} & \colhead{Value}}
\startdata
$\mathrm{wabs}$  & $N_{\rm H}$ $(\times10^{22}\,\mathrm{cm^{-2}})$           &  $0.68\pm0.44$ \\
\hline
\multirow{6}{*}{$\mathrm{vrnei}$} & $\mathrm{Norm.}$ $(\times10^5)$          &  $2.32\pm7.13$ \\
& $kT$ $\mathrm{(keV)}$                                                    &  $1.46\pm0.27$ \\
& $kT_{init}$ $\mathrm{(keV)}$                                             &  $0.31\pm0.63$ \\
& $\mathrm{\tau}$ $\mathrm{(\times10^{13}\,cm^{-3}\,s)}$                   &  $>4.53$ \\
& $\mathrm{Mg}$ $(Z_\odot)$                              &  $6.11\pm5.12$ \\
& $\mathrm{Si}$ $(Z_\odot)$                              &  $28.6\pm28.4$ \\
\hline
\multirow{4}{*}{$\mathrm{vmekal}$} & $\mathrm{Norm.}$ $(\times10^5)$         &  $4.35\pm0.68$ \\
&$kT$ $\mathrm{(keV)}$                                            &  $10.1\pm11.4$ \\
&$\mathrm{Mg}$ $(Z_\odot)$                               &  $\leq9.67$ \\
&$\mathrm{Si}$ $(Z_\odot)$                               &  $\leq4.61$ \\
\enddata
\tablecomments{Parameters are: $N_{\rm H}$, the column density of the neutral Hydrogen; $kT$, the electron temperature; $kT_{init}$, the initial plasma temperature; $\tau$, the ionization timescale; and the abundances of each element.}
\end{deluxetable}

Furthermore, we employ the non-linear $L_\mathrm{X}-\mathrm{SFR}$ correlation derived by \citet{2003MNRAS.339..793G}
\begin{equation}
    \mathrm{SFR\,(\mathrm{M_\odot\,yr^{-1}})} = \left(\frac{L_\mathrm{X,2-10keV}}{2.6\times10^{39}\,\mathrm{erg\,s^{-1}}}\right)^{0.6}, 
\end{equation}
to estimate the X-ray emissions produced by star-forming activities. This non-linear correlation only works for low SFR galaxies \citep[e.g.][]{2014MNRAS.437.1698M}. Taking the $\mathrm{SFR=3.2\,M_\odot\,yr^{-1}}$ \citep[or $\mathrm{SFR=4.8\,M_\odot\,yr^{-1}}$, see][]{2015ApJ...815..133S} estimated from $\mathrm{IR}$ luminosity \citep{2012ApJ...752...38T}, the X-ray luminosity of \obj\ is calculated as $\log{(L_\mathrm{X,2-10keV}/\mathrm{erg\,s^{-1}})}=40.25$, while the $Chandra$ observations obtain a intrinsic (i.e. unabsorbed) X-ray luminosity $\log{(L_\mathrm{X,2-10keV}/\mathrm{erg\,s^{-1}})}=39.03\pm0.05$. The estimated $\mathrm{2-10\,keV}$ X-ray luminosity via the above equation is $\sim1.22$-dex higher than the observations. Taking the recent $L_\mathrm{X}-\mathrm{SFR}$ correlation derived by \citet{2020NatAs...4..159B} still results in an overestimation of X-ray luminosity.

These results imply insufficient XRBs in the central region of the galaxy \obj, and they disfavor an AGN as well. Indeed, the high resolution ($\sim0.5^{\prime\prime}$) \telescope{Chandra} observation shows that the X-ray emission neither overlaps with the radio emissions (see panel c of Figure \ref{fig:full}) nor associates with the individual radio source \citep[e.g. the cases of X-ray SNe/SNRs,][]{2014ApJS..212...21L}. Furthermore, the X-ray emission has a symmetric morphology extending in both the north and south of the galaxy disk, which mimics a galactic scale wind \citep[see also][]{2012ApJ...752...38T}. Interestingly, the \telescope{Chandra} spectrum presented in this work (including both diffuse and point-like X-ray emission) is highly consistent with the point-source subtracted spectrum \citep[see][]{2012ApJ...752...38T}, which again implies a deficit of point X-ray sources.

\begin{figure}
\centering 
\includegraphics[width=0.47\textwidth]{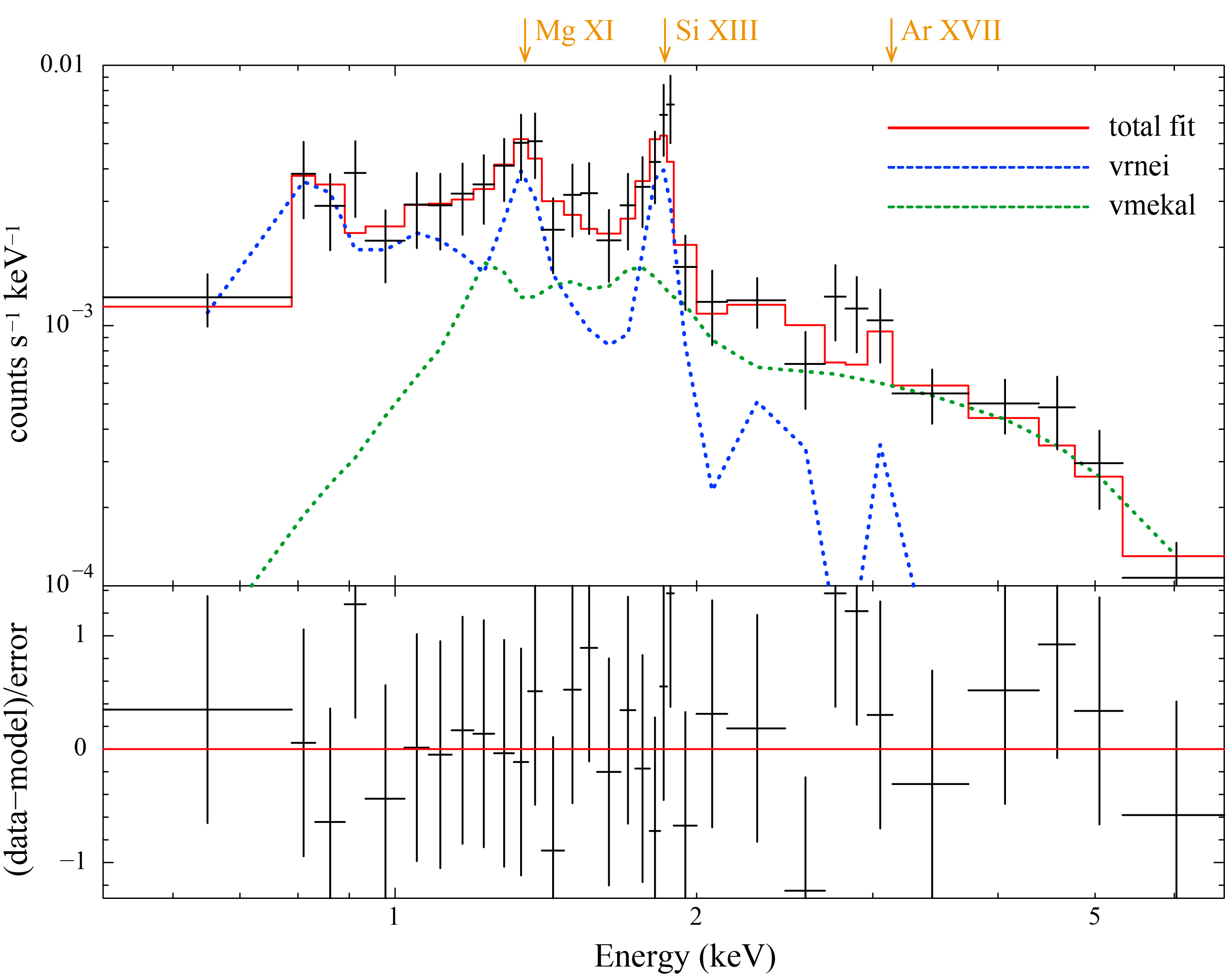}
\caption{\telescope{Chandra} $0.5-7$\,keV X-ray spectrum of \obj. The red solid line corresponds to the best-fitting model \module{wabs*(vrnei+vmekal)}, where the \module{vrnei} model shows in blue dashed line and the \module{vmekal} model shows in green dashed line. The residuals of the best fit are shown in the bottom panel. \label{fig:line}}
\end{figure}

Therefore, we account for the X-ray emissions from activities that are triggered by SNRs. We modeled the \telescope{Chandra} spectra with a combination of a thermal gas (\module{vmekal} model in \softw{xspec}) and a nonequilibrium ionization plasma (\module{vrnei} model in \softw{xspec}), which correspond to the interstellar medium and the ejecta of SNRs \citep[see][]{2020PASJ...72...65S}, respectively. The abundance of elements O, Fe, Ne, Mg, Si, and Ar are free parameters, while the others are fixed to the solar abundance. The best-fit parameters and spectral lines are presented in Table \ref{tab:fit_vv} and Figure \ref{fig:line}, respectively. The fitting yields electron temperatures of $1.46\pm0.27\,\mathrm{keV}$ and $10.1\pm11.4\,\mathrm{keV}$ for NEI and MeKaL medium, respectively. Interestingly, the fit not only represents the emission lines Mg XI and Si\,\textsc{xiii}, but also the lines Ar\,\textsc{xvii} at 3.104\,keV and the combination of O\,\textsc{viii}, Fe\,\textsc{xvii}, and Ne\,\textsc{ix} lines at 0.7$-$0.8\,keV. The abundances of Mg and Si are roughly consistent with the \module{wabs*(powerlaw+vapec)} and \module{wabs*(powerlaw+vmekal)}. Obviously, this model better represents the spectrum and the associated narrow features than the above models, though it has an over-fitting $\chi^{2}_\mathrm{red}=0.62$ due to the limited data points. As a result, this model indicates that the X-ray emissions in the central region of \obj\ can be fully explained by SN-triggered activities. Along with the morphology of the X-ray emitting regions, the X-ray emissions trace a hot and highly ionized galactic scale gas outflow of \obj.

\section{Conclusion}
We conducted a study on the radio and X-ray emissions of the starburst galaxy \obj, and have summarized our findings as follows. (1) Through the VLBA 1.5\,GHz observation, we achieved the highest resolution of $\sim5-10$\,mas and were able to resolve the central dominating radio sources of this galaxy. This ensues the identification of seven compact SNRs in galaxy \obj, and three of them are consistent with partial shells; (2) We estimated the diameter of these seven sources by fitting an optically thin sphere model, and we discovered that one of these seven sources is a new radio SN; (3) We explored the $\Sigma-D$ relation by using the SNRs detected in \obj\ and other SNR samples with accurate size estimations; (4) We analyzed archival observations of \obj\ from \telescope{Chandra} and \telescope{XMM-Newton} with a total effective exposure time of more than 170 ks. We found that the X-ray spectra can be well modeled with the activities of SNRs; (5) We investigated the correlation between X-ray luminosity and star-formation rate, as well as the correlation between radio and X-ray luminosity for starburst galaxies. Our findings suggest that the X-ray emission is more likely to originate from the activities of SNRs than from X-ray binaries and AGNs. These results indicate that SN-triggered activities are responsible for producing both X-ray and radio emissions in the starburst galaxy \obj. This work provides valuable insights into the SN-driven radio activity and galactic-scale gas outflow model.

\begin{acknowledgments}
This research was supported by the National Key R\&D Program of China (Grant No. 2020YFE0202100), the Shanghai Sailing Program (21YF1455300), the National Science Foundation of China (12103076), and the China Postdoctoral Science Foundation (2021M693267). Scientific results from the data presented in this publication are derived from the VLBA Project BA146. The National Radio Astronomy Observatory is a facility of the National Science Foundation operated under cooperative agreement by Associated Universities, Inc.
\end{acknowledgments}

\clearpage
\bibliography{ngc3628}{}
\bibliographystyle{aasjournal}



\end{CJK*}
\end{document}